\documentclass[11pt]{article}

\usepackage[final]{acl}

\usepackage{times}
\usepackage{latexsym}

\usepackage[T1]{fontenc}

\usepackage[utf8]{inputenc}

\usepackage{microtype}
\usepackage{inconsolata}
\usepackage{graphicx}
\usepackage{amsmath}
\DeclareMathOperator*{\argmax}{arg\,max}
\usepackage{amssymb}
\usepackage{dsfont}
\usepackage{booktabs}
\usepackage{algorithm}
\usepackage{algorithmic}
\usepackage{multirow}
\usepackage{enumitem}
\setlist{nosep}
\usepackage{tcolorbox}
\tcbuselibrary{breakable,skins}
\usepackage{tabularx}
\usepackage{stfloats}

\newtcolorbox{promptbox}[1]{
    colback=gray!5,     % Light gray background
    colframe=black!75,  % Dark gray border
    fonttitle=\bfseries\small,
    title=#1,           % The title of the prompt
    arc=2mm,            % Rounded corners
    boxrule=0.5pt,      % Border thickness
    left=5pt, right=5pt, top=5pt, bottom=5pt,
    fontupper=\small\ttfamily, % Monospace font for the content
    breakable           % Allows box to span multiple pages
}

\title{Beyond the Vacuum: Combinatorial Strategy Selection for Competitor-Aware Generative Engine Optimization}

 \author{
    Vaibhav Sourirajan \quad Yao Zhang \quad Himanshu Kumar \quad Sahil Wadhwa \\
    \textbf{Mann Patel} \quad \textbf{Amirfarrokh Iranitalab} \\
    Capital One, AI Foundations \\
    \texttt{\{vaibhav.sourirajan, yao.zhang, himanshu.kumar2, sahil.wadhwa,} \\
    \texttt{mann.patel2, amirfarrokh.iranitalab\}@capitalone.com}
  }

\begin{document}
\maketitle

\begin{abstract}
Generative Engine Optimization (GEO) has emerged as a novel paradigm for transforming content to increase visibility in Large Language Model (LLM) responses. Traditional GEO methods, however, select rewriting strategies in isolation, ignoring a critical externality: as adoption of content optimization grows, optimal strategies for rewriting content change. We formalize GEO as a \emph{competitor-aware} strategy selection problem and propose a two-phase pipeline to solve it: (1) We use Bayesian Optimization of Combinatorial Structures (BOCS) to efficiently search the space of rewriting strategies, (2) We generate preference pairs and grounded reasoning traces from the BOCS black-box observations to fine-tune a language model to analyze a document corpus and propose optimal rewriting strategy combinations. We achieve state-of-the-art performance across several impression metrics over existing agentic and single-heuristic methods on both \texttt{geo-bench} and our synthetically augmented competitive dataset $\texttt{geo-bench}_{comp}$. Our method also transfers to multiple out-of-distribution datasets, proving effective across domains, queries, and document types.

\end{abstract}

\section{Introduction}
\label{sec:intro}

The way users access information on the web is undergoing a major shift. Traditional search engines are rapidly being augmented or replaced by generative answer engines that leverage LLMs to synthesize direct, conversational responses \cite{chen2026navigating}. Consequently, content creators are shifting focus from traditional Search Engine Optimization (SEO) \cite{davis2006search, almukhtar2021search} to Generative Engine Optimization (GEO): the practice of rewriting source content to maximize visibility and citation likelihood in LLM-generated responses.

Early GEO work showed that single rewriting heuristics (eg. adding quotations, authoritative tone) and agentic methods that identify broad content improvement trends can significantly boost a document's representation in generative outputs \cite{aggarwal2024geo, autogeo, mageo, sageoarena, kumar2024manipulating}. However, these methods optimize a target document in isolation, producing fixed rewrites without awareness of the surrounding competitive corpus. When several documents apply the same rewriting strategies, overall gains decrease, revealing a zero-sum nature to content optimization \cite{cseobench}. Effective GEO is thus a relative problem, requiring methods that consider the full competitive landscape.

Motivated by these insights, we formalize GEO as a \textit{competitor-aware} strategy selection problem: given a query, a target document, and the surrounding corpus, we seek to find the combination of strategies maximizing the target document's visibility in generative engine responses. We implement a two-stage pipeline for training a selector language model to propose optimal, reasoning-backed rewriting strategies. The first stage employs Bayesian Optimization of Combinatorial Structures (BOCS) \cite{baptista2018bocs} to efficiently search the strategy space, collecting black-box observations of downstream visibility for various rewriting combinations. The second stage curates a training dataset from noisy BOCS observations by introducing a statistically principled hard-negative mining algorithm that uses Hamming distance to surface preference pairs. To ensure the selector robustly maps document context to rewriting strategies, we generate contrastive post-hoc reasoning traces using a teacher LLM. The selector is fine-tuned via Direct Preference Optimization (DPO) \cite{rafailov2023dpo} on grounded reasoning and strategy pairs, enabling the model to internalize relative document dynamics and identify comparative advantages. We perform evaluation on \texttt{geo-bench} and $\texttt{geo-bench}_{comp}$, a synthetically augmented version of \texttt{geo-bench} which introduces competitor document optimization at varying adoption rates. Our approach outperforms 15 single-strategy and 2 agentic baselines across several citation and LLM-based metrics on both datasets. Crucially, on $\texttt{geo-bench}_{comp}$, our method is resilient to scaling adoption rate, exhibiting the slowest rate of performance degradation. Further analysis demonstrates that grounding strategy combinations with teacher-generated reasoning traces unlocks strong generalization capabilities, mitigating hallucinated analysis and surface-level pattern recognition. Our main contributions are as follows:
\begin{enumerate}
    \item We are the first to formalize, implement, and evaluate competitor-aware GEO. To support this, we introduce $\texttt{geo-bench}_{comp}$, a competitive benchmark that enables the evaluation of optimization strategies under varying GEO adoption rates.
    \item We design a two-stage data curation pipeline combining BOCS-driven combinatorial search, statistically grounded hard-negative mining, and teacher-generated reasoning traces to produce high-quality DPO training pairs.
    \item We fine-tune a selector model that achieves state-of-the-art performance on both \texttt{geo-bench} and $\texttt{geo-bench}_{comp}$, as well as robust transferability to out-of-distribution datasets.
\end{enumerate}

\begin{figure*}[t]
    \centering
    \includegraphics[width=\textwidth]{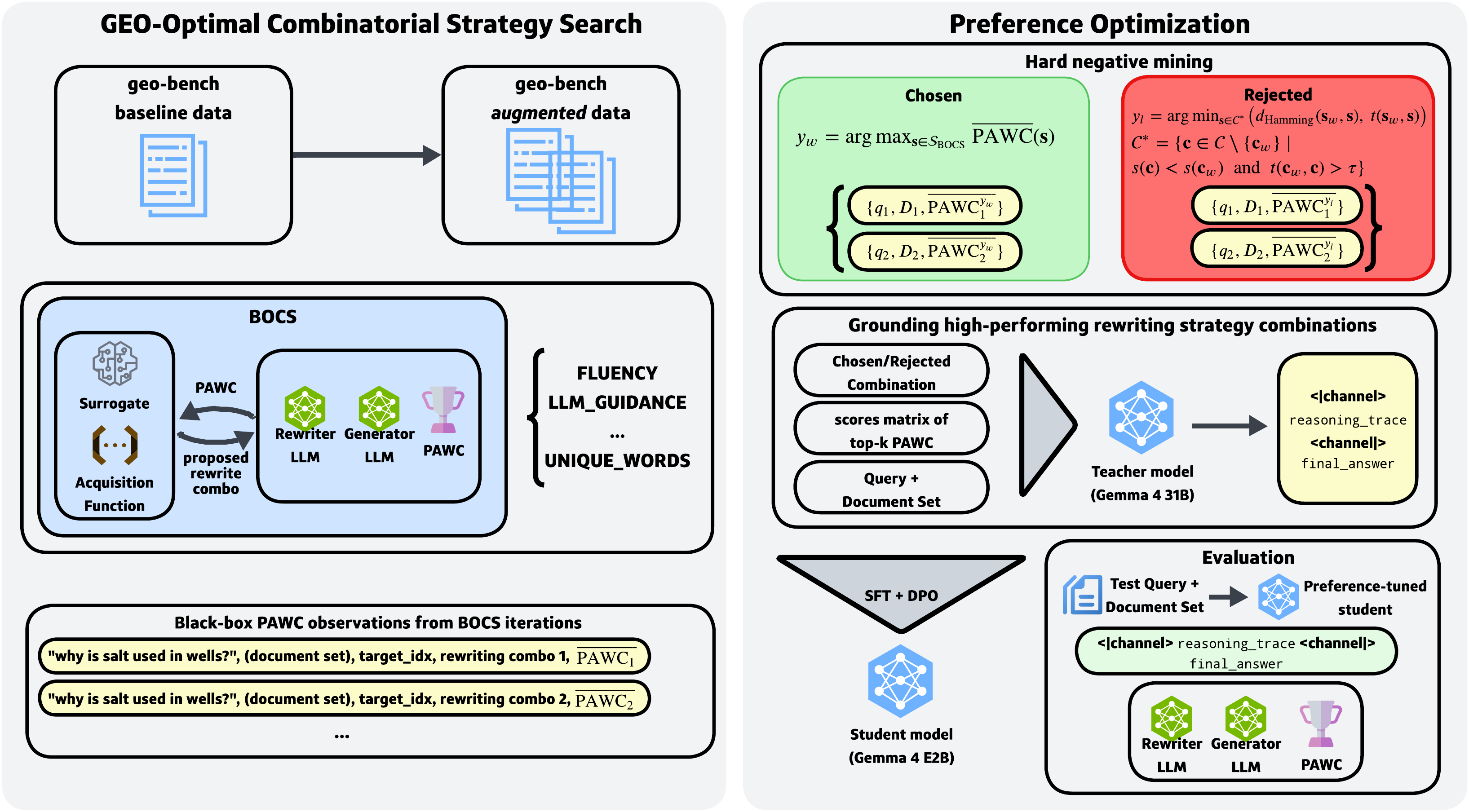}
    \caption{Overview of the Competitor-Aware GEO pipeline. \textbf{Phase 1 (GEO-Optimal Combinatorial Strategy Search):} We produce $\texttt{geo-bench}_{comp}$ through competitor simulation, drawing rewriting strategies from a mixed distribution for varying adoption rates $\alpha$. For these datasets, BOCS searches the $2^{15}$ strategy space per (query, document set) pair, evaluating candidates via a rewrite-once/evaluate-five black-box function. \textbf{Phase 2 (Preference Optimization):} Hard negative mining extracts statistically significant preference pairs from BOCS observations; a teacher model generates contrastive reasoning traces for each pair; a selector language model is then fine-tuned via a two step SFT+DPO recipe on the resulting preference pairs. At evaluation time, the fine-tuned selector receives only the query and document corpus and outputs a reasoning-backed strategy combination, which is applied to the target document and evaluated for downstream visibility.}
    \label{fig:pipeline}
\end{figure*}

\section{Related Work}
\label{sec:related}

The concept of GEO was formalized by \citet{aggarwal2024geo}, which also introduced the \texttt{geo-bench} dataset and demonstrated that targeted rewriting strategies, such as adding quotations, could significantly boost a document's visibility in LLM-generated responses. This finding sparked multiple lines of inquiry, including automating strategy discovery and formalizing robust GEO evaluation systems. AutoGEO \cite{autogeo} tackled strategy discovery by extracting optimization rules that caused the largest changes in citation rates, merging extracted heuristics into concrete, non-overlapping rewriting strategies. As the field matured, evaluation frameworks extended beyond word count metrics: CC-GSEO-Bench \cite{ccgseobench} introduced an LLM-as-a-judge approach to measure a document's marginal contribution by comparing responses with and without it, while C-SEO Bench \cite{cseobench} measured delta-rank ($\Delta rank)$ of citation position. \citet{cseobench} took a critical step further by introducing the notion of adoption rate to quantify the fraction of documents adopting GEO strategy. Importantly, they found that strategy effectiveness degrades as more competitors optimize document content, directly motivating our work: if strategies lose marginal gains as adoption grows, then optimizing in isolation is fundamentally flawed.

Building on these foundations, subsequent work pushed GEO toward more realistic settings and sophisticated optimization. SAGEO Arena \cite{sageoarena} introduced a benchmark requiring documents to survive a full retrieval and re-ranking pipeline before generation, revealing that strategies effective in isolation often fail under real-world retrieval conditions. They additionally preserve key structural elements in web documents, indexing meta descriptions and schema markup of scraped web pages, enabling realistic evaluation of strategies. On the optimization side, Product Bench \cite{productbench} proposed controlling output rankings by directly targeting document embedding representations, while MAGEO \cite{mageo} developed a multi-agent framework that learns and reuses optimization strategies across documents. AgenticGEO \cite{agenticgeo} pushed the frontier, introducing a content-conditioned methodology which utilized a continuous Evolver-Critic setup for multi-turn rewriting. Despite these advances in GEO study, current methodologies treat document optimization as an isolated generative task. By ignoring the surrounding competitive corpus, they fail to adapt to the shifting baseline created when multiple documents employ GEO strategies. We argue that robust content optimization must move beyond static heuristics and isolated agentic methods in favor of conditioning on the full competitive environment.

% \paragraph{Combinatorial Optimization and Preference Learning.}
% Closing this gap requires three design choices that simpler alternatives cannot satisfy. First, the $2^{15}$ strategy space cannot be navigated by grid search or random sampling at reasonable cost: we require an efficient surrogate-based optimizer, motivating our use of BOCS \cite{baptista2018bocs}, which has proven effective for binary combinatorial spaces but has never been applied to content optimization. Second, the noisy black-box observations BOCS produces cannot be directly converted into reliable training pairs: naive pairing of high and low scoring combinations conflates true strategy differences with evaluation variance, making supervised fine-tuning on such pairs unreliable. This motivates our statistically principled hard-negative mining approach and the use of DPO \cite{rafailov2023dpo} over SFT, since DPO's contrastive objective is explicitly designed to learn from relative preferences rather than absolute labels. Third, a selector trained only on strategy labels without reasoning will fail to generalize when the competitive landscape shifts: the Orca line of work \cite{mukherjee2023orca, mitra2023orca2} demonstrated that distilling chain-of-thought reasoning from a strong teacher substantially improves student generalization, a finding \citet{ma2025cotdpo} extended specifically to DPO. We are the first to bring these tools together in service of competitor-aware GEO, and each choice is driven by a concrete limitation of the available alternatives.

\section{Problem Formulation}
\label{sec:formulation}

We formalize competitor-aware GEO as a content-conditioned combinatorial optimization problem: given a query, a target document, and a surrounding competitive corpus, select the rewriting strategy combination maximizing the target document's visibility. Let $q$ denote a query, $\mathcal{D} = \{d_1, \ldots, d_k\}$ an initial document corpus for $q$, and index $j \in \{1, \ldots, k\}$ identifying the document to optimize. Given an adoption rate $\alpha \in [0, \frac{k-1}{k}]$, we construct the competitive corpus $\mathcal{D}_\alpha$ by rewriting $\alpha$ fraction of total documents (excluding the target document) according to the competitor simulation (\S\ref{sec:competitor}). Note, in practice, the $\alpha$ space is discretized by the number of documents, taking on $k$ distinct values. A rewriting strategy combination $\mathbf{x} \in \{0,1\}^{r}$, where each dimension $x_i$ represents the the inclusion or exclusion of heuristic $i$ in the rewriting prompt, transforms the target document into $d'_j=d_j(\mathbf{x})$, producing the final evaluation corpus $\mathcal{D}'_\alpha$ by substituting the rewritten target into the competitive corpus. A generative engine $\mathcal{G}$ receives the query and full document corpus and produces a cited response:
\begin{equation}
    R = \mathcal{G}(q, \mathcal{D}'_\alpha(\mathbf{x}))
\end{equation}
We seek the strategy combination that maximizes the impression of the target source in the generated response:
\begin{equation}
    \mathbf{x}^* = \argmax_{\mathbf{x} \in \{0,1\}^{r}} \; \text{Impression}(j, R)
\end{equation}
Following \citet{aggarwal2024geo}, we utilize Position-Adjusted Word Count (PAWC), an exponentially decaying weighted citation function, as the primary impression metric for evaluation. Let $R$ consist of sentences $\{s_1, \ldots, s_N\}$, and let $\text{cites}(s_i)$ denote the multiset of citation indices extracted from sentence $s_i$. The PAWC score for target source index $j$ is:
\begin{equation}
    \text{PAWC}(j, R) = \sum_{i=1}^{N} \frac{c_j(s_i) \cdot |s_i| \cdot w_i}{|\text{cites}(s_i)|}
\end{equation}
where $c_j(s_i) = \sum_{c \in \text{cites}(s_i)} \mathds{1}[c = j]$ counts citations to source $j$ in sentence $s_i$, $|s_i|$ counts words with more than 2 characters, and $w_i = \exp(-i/(N{-}1))$ is an exponential position decay giving earlier sentences higher weight.

\subsection{Competitor Simulation}
\label{sec:competitor}
To construct $\mathcal{D}_\alpha$, we model a heterogeneous competitive landscape. We define three competitor rewriting functions:
\begin{enumerate}
    \item $\mathcal{W}_1$ (\textbf{Random Combo}): Rewrite using a uniformly sampled $\mathbf{x} \sim \text{Uniform}(\{0,1\}^{r})$
    \item $\mathcal{W}_2$ (\textbf{Strongest Single}): Rewrite using the empirically strongest individual strategy (Quotation Addition, $\mathbf{e}_4$)
    \item $\mathcal{W}_3$ (\textbf{AutoGEO}): Rewrite using the full set of extracted heuristics from \citet{autogeo}
\end{enumerate}

For a given $\alpha$, we compute the number of documents to rewrite as $n_\alpha = \alpha |\mathcal{D}| = \alpha k$. We then sample $n_\alpha$ indices without replacement from the competitor document set $\mathcal{D} \setminus \{d_j\}$ per data point $i$. The selected indices are sorted in ascending order and each is assigned a rewriting function $w \in \{\mathcal{W}_1, \mathcal{W}_2, \mathcal{W}_3\}$ via fixed cyclic allocation. All non-selected documents remain unchanged. This produces a heterogeneous competitive corpus where, for any $n_\alpha \geq 3$, all three competitor types are represented in roughly equal proportion. The deterministic seeding ensures reproducibility, while the per-$\alpha$ independent sampling avoids artificial nesting across adoption rates.

\begin{algorithm}[t]
\caption{Hard Negative Mining for GEO}
\label{alg:hard_neg}
\begin{algorithmic}[1]
\REQUIRE BOCS observations $\{(\mathbf{x}_i, \mu_i, \text{SE}_i)\}_{i=1}^{N}$ for a single training point, each $\mathbf{x}_i \in \{0,1\}^{15}$ is a strategy combination evaluated during BOCS, $\mu_i$, $\text{SE}_i$ are the sample mean and standard error; $(\mu_0, \text{SE}_0)$ are the mean and standard error of the zero-vector baseline; threshold $t_{\text{thresh}} = 1.860$
\ENSURE A preference pair $(\mathbf{x}^+, \mathbf{x}^-)$---the chosen and rejected strategy combinations---or \textsc{None} if no valid pair exists
\STATE $\mathbf{x}^+ \leftarrow \argmax_i \mu_i$
\STATE $(\mu_c, \text{SE}_c) \leftarrow (\mu_{x^+}, \text{SE}_{x^+})$
\IF{$(\mu_c - \mu_0) / \sqrt{\text{SE}_c^2 + \text{SE}_0^2} \leq t_{\text{thresh}}$}
    \RETURN \textsc{None} \COMMENT{Chosen does not significantly beat baseline}
\ENDIF
\STATE $\mathcal{R} \leftarrow \emptyset$
\FOR{each $(\mathbf{x}_i, \mu_i, \text{SE}_i)$ with $\mu_i < \mu_c$}
    \STATE $t_i \leftarrow (\mu_c - \mu_i) / \sqrt{\text{SE}_c^2 + \text{SE}_i^2}$
    \IF{$t_i > t_{\text{thresh}}$}
        \STATE $h_i \leftarrow \text{Hamming}(\mathbf{x}^+, \mathbf{x}_i)$
        \STATE $\mathcal{R} \leftarrow \mathcal{R} \cup \{(\mathbf{x}_i, h_i, t_i)\}$
    \ENDIF
\ENDFOR
\STATE Sort $\mathcal{R}$ by $(h_i \uparrow, t_i \uparrow)$ \COMMENT{Closest to chosen (min Hamming), tie-broken by smallest significant $t$ (hardest)}
\RETURN $(\mathbf{x}^+, \mathcal{R}[0])$
\end{algorithmic}
\end{algorithm}

\section{Methodology}
\label{sec:method}

\subsection{Overview}

Our methodology has two phases (Figure~\ref{fig:pipeline}): (1)~GEO-optimal combinatorial strategy search via BOCS, and (2)~preference optimization of a selector LLM via DPO.

\subsection{GEO-Optimal Combinatorial Strategy Search}
\label{sec:data_collection}

We start from the \texttt{geo-bench} training split \citep{aggarwal2024geo}, containing 7{,}998 queries, each with 5 documents, spanning various domains and query types. We augment this into $\texttt{geo-bench}_{\text{comp}}$ by constructing competitive corpora at each $\alpha \in \{0.0, 0.2, 0.4, 0.6, 0.8\}$ according to the mixed competitor distribution defined in \S\ref{sec:competitor}, yielding 39{,}990 total training examples. 

For evaluation, we similarly augment the  \texttt{geo-bench} test split (1000 samples) following the same mixed competitor distribution to produce a $\texttt{geo-bench}_{\text{comp}}$ test split of size 5000. For each data point in \texttt{geo-bench} and $\texttt{geo-bench}_{\text{comp}}$, we run Bayesian Optimization of Combinatorial Structures (BOCS) \cite{baptista2018bocs} to efficiently search the strategy space, collecting black-box observations from which statistically grounded preference pairs are later mined.

\paragraph{Strategy combination space.}
We define 15 binary rewriting strategies yielding a search space of size $2^{15} = 32{,}768$ possible combinations: Fluency, Unique Words, Authoritative, Quotation Addition, Citing Sources, Easy Understanding, Technical Terms, Keyword Stuffing, and Statistics Addition from \citet{aggarwal2024geo} (rewriting prompts adopted from AgenticGEO \cite{agenticgeo}); LLM Guidance from C-SEO Bench \cite{cseobench}; and five additional strategies introduced in this work---Expanding Facts, Structured Formatting, Balanced View, Conciseness, and Pros and Cons---which formalize heuristics discovered by AutoGEO \cite{autogeo}. We include all single-strategy rewriting prompts and composed rewriting prompt for multi-strategy rewriting in Appendix~\ref{sec:appendix_prompts}.

\subsubsection{BOCS Training}

Following \citet{baptista2018bocs}, given an expensive-to-evaluate black-box function $f$ over domain $D = \{0,1\}^d$, we seek to find $\arg \max_{x\in D}f(x)$. In our case, $D = \{0,1\}^{15}$.

\paragraph{Surrogate model and acquisition function.}
We use a second-order polynomial with $1 + d + \binom{d}{2} = 121$ terms---a constant, $d{=}15$ linear terms, and $\binom{d}{2} = 105$ pairwise interaction terms:
\begin{equation}
\begin{aligned}
    f(\mathbf{x}) &= \beta_0 + \sum_{j=1}^{d} \beta_j x_j + \sum_{j<k} \beta_{jk} x_j x_k 
\end{aligned}
\end{equation}

We place a horseshoe prior \citep{carvalho2010horseshoe} over the coefficients $\boldsymbol{\beta}$. We use Gibbs sampling and leverage fast multivariate Gaussian samplers \citep{bhattacharya2016fast,rue2001fast} for posterior inference. Given a posterior sample $\tilde{\boldsymbol{\beta}}$, we use simulated annealing \citep{bertsimas1993simulated} as the acquisition function to optimize the surrogate prediction over $\{0,1\}^d$. We run BOCS for a total of 50 function evaluations (10 random initializations followed by up to 40 surrogate-guided iterations), with early stopping patience. We provide the full algorithm hyperparameters in Appendix~\ref{sec:appendix_bocs}.

\paragraph{Black-box evaluation.}
Each evaluation of a candidate combination $\mathbf{x}$ proceeds in two steps. First, the target document $d_j$ is rewritten once at temperature 0 using the composed strategy prompt for $\mathbf{x}$ (Appendix~\ref{sec:appendix_composition}), producing $d_j(\mathbf{x})$. Second, the generative engine evaluates the resulting corpus $n{=}5$ independent times at temperature 0.7, yielding PAWC scores $\{\text{PAWC}_1, \ldots, \text{PAWC}_5\}$. We record the sample mean $\mu(\mathbf{x})$ and standard error $\text{SE}(\mathbf{x})$.

\begin{table*}[!t]
\centering
\fontsize{8.5}{9}\selectfont
\setlength{\tabcolsep}{3pt}
\begin{tabular}{lccccccc}
\toprule
& \multicolumn{3}{c}{\textbf{gpt-oss-120b}} & & \multicolumn{3}{c}{\textbf{Llama-3.3-70B-Instruct}} \\
\cmidrule(lr){2-4} \cmidrule(lr){6-8}
\textbf{Method} & \textbf{PAWC} & \textbf{Pos Count} & \textbf{Word Count} & & \textbf{PAWC} & \textbf{Pos Count} & \textbf{Word Count} \\
\midrule
No Rewrite & $20.03 \pm 0.24$ & $20.02 \pm 0.33$ & $19.99 \pm 0.23$ & & $20.18 \pm 0.20$ & $20.13 \pm 0.20$ & $20.14 \pm 0.17$ \\
\midrule
Fluency Optimization & $21.92 \pm 0.32$ & $21.79 \pm 0.37$ & $21.77 \pm 0.22$ & & $21.42 \pm 0.14$ & $21.48 \pm 0.15$ & $21.37 \pm 0.16$ \\
Unique Words & $20.22 \pm 0.29$ & $20.23 \pm 0.28$ & $20.17 \pm 0.26$ & & $19.52 \pm 0.20$ & $19.66 \pm 0.22$ & $19.54 \pm 0.21$ \\
Authoritative & $21.02 \pm 0.17$ & $20.91 \pm 0.18$ & $20.99 \pm 0.21$ & & $20.97 \pm 0.14$ & $21.10 \pm 0.11$ & $20.98 \pm 0.22$ \\
Quotation Addition & $21.96 \pm 0.33$ & $21.95 \pm 0.25$ & $21.86 \pm 0.22$ & & $22.59 \pm 0.13$ & $22.76 \pm 0.15$ & $22.54 \pm 0.14$ \\
Citing Sources & $21.72 \pm 0.30$ & $21.60 \pm 0.35$ & $21.53 \pm 0.26$ & & $21.42 \pm 0.08$ & $21.40 \pm 0.13$ & $21.49 \pm 0.08$ \\
Easy Understanding & $22.03 \pm 0.20$ & $21.95 \pm 0.23$ & $21.95 \pm 0.20$ & & $21.34 \pm 0.10$ & $21.38 \pm 0.13$ & $21.35 \pm 0.07$ \\
Technical Terms & $21.40 \pm 0.15$ & $21.29 \pm 0.15$ & $21.33 \pm 0.17$ & & $19.52 \pm 0.06$ & $19.61 \pm 0.05$ & $19.67 \pm 0.08$ \\
Keyword Stuffing & $20.52 \pm 0.10$ & $20.53 \pm 0.09$ & $20.58 \pm 0.09$ & & $21.67 \pm 0.30$ & $21.78 \pm 0.27$ & $21.76 \pm 0.33$ \\
Statistics Addition & $23.08 \pm 0.28$ & $23.14 \pm 0.25$ & $22.91 \pm 0.24$ & & $22.38 \pm 0.16$ & $22.42 \pm 0.22$ & $22.44 \pm 0.13$ \\
All In One & $22.40 \pm 0.44$ & $22.35 \pm 0.44$ & $22.17 \pm 0.38$ & & $21.83 \pm 0.14$ & $21.93 \pm 0.15$ & $21.83 \pm 0.08$ \\
LLM Guidance & $23.18 \pm 0.19$ & $23.30 \pm 0.21$ & $23.27 \pm 0.20$ & & $22.36 \pm 0.19$ & $22.62 \pm 0.17$ & $22.39 \pm 0.19$ \\
Expanding Facts & $22.16 \pm 0.21$ & $22.09 \pm 0.13$ & $21.92 \pm 0.24$ & & $22.55 \pm 0.24$ & $22.42 \pm 0.22$ & $22.31 \pm 0.32$ \\
Structured Formatting & $23.04 \pm 0.43$ & $23.12 \pm 0.36$ & $22.94 \pm 0.40$ & & $21.21 \pm 0.18$ & $21.40 \pm 0.16$ & $21.34 \pm 0.14$ \\
Balanced View & $19.44 \pm 0.24$ & $19.49 \pm 0.25$ & $19.77 \pm 0.27$ & & $20.95 \pm 0.11$ & $20.95 \pm 0.07$ & $21.12 \pm 0.16$ \\
Conciseness & $23.44 \pm 0.22$ & $23.50 \pm 0.21$ & $23.48 \pm 0.19$ & & $21.07 \pm 0.31$ & $21.37 \pm 0.24$ & $21.12 \pm 0.28$ \\
Pros and Cons & $20.71 \pm 0.21$ & $20.63 \pm 0.18$ & $20.84 \pm 0.19$ & & $22.47 \pm 0.49$ & $22.57 \pm 0.49$ & $22.69 \pm 0.46$ \\
\midrule
AutoGEO & $27.16 \pm 0.44$ & $27.16 \pm 0.46$ & $26.93 \pm 0.41$ & & $23.85 \pm 0.24$ & $23.73 \pm 0.22$ & $24.05 \pm 0.28$ \\
AgenticGEO & $27.95 \pm 0.30$ & $27.72 \pm 0.29$ & $28.06 \pm 0.22$ & & $26.80 \pm 0.08$ & $26.38 \pm 0.11$ & $27.04 \pm 0.07$ \\
\midrule
Competitor-Aware GEO (Ours) & $\mathbf{32.62 \pm 0.25}$ & $\mathbf{32.86 \pm 0.28}$ & $\mathbf{32.36 \pm 0.24}$ & & $\mathbf{29.55 \pm 0.15}$ & $\mathbf{29.85 \pm 0.10}$ & $\mathbf{29.14 \pm 0.08}$ \\
\bottomrule
\end{tabular}
\caption{Overall performance on \texttt{geo-bench} (original corpus, no competitor rewrites). We report mean and standard deviation across 1 rewrite (temperature=0) and 5 evaluations from the generative engine (temperature=0.7).}
\label{tab:results_original}
\end{table*}

\subsection{Preference Optimization}
\label{sec:dpo}

Preference optimization proceeds in three phases: (1) hard negative mining to identify high-quality contrastive pairs, (2) reasoning trace generation to ground each pair in empirical BOCS evidence, and (3) preference tuning of the selector model.

\subsubsection{Hard Negative Mining}
\label{sec:hard_neg}
We employ a hard negative mining strategy (Algorithm~\ref{alg:hard_neg}) that operates independently on each query--document set pair, producing a single preference pair $(\mathbf{x}^+, \mathbf{x}^-)$ from all black-box BOCS observations collected for that point. We select the chosen combination $\mathbf{x}^+$ as the BOCS observation with the highest mean PAWC, provided it statistically significantly outperforms the zero-vector baseline under a one-tailed Welch's $t$-test with threshold $t_{\text{thresh}} = 1.860$. If no such combination exists, the training point is discarded. Candidate rejected combinations are then drawn from observations with lower mean PAWC than $\mathbf{x}^+$ that are also statistically significantly worse under the same test. Among these, we select the \emph{hardest} negative, which minimizes Hamming distance to $\mathbf{x}^+$, breaking ties by the smallest statistically significant $t$-statistic. This ensures $\mathbf{x}^-$ is maximally similar to $\mathbf{x}^+$ in strategy space yet reliably distinguishable in performance.

\subsubsection{Score-Grounded Reasoning Trace Generation}
\label{sec:cot}

The preference pairs produced by hard negative mining are strategy vectors $(\mathbf{x}^+, \mathbf{x}^-)$. Without intermediate reasoning, DPO over short strategy lists has no rich signal to learn from \citep{liu2025uncovering}. Following Orca-style distillation \citep{mukherjee2023orca, mitra2023orca2}, we independently generate reasoning traces leading to the final strategy combination using a teacher model: \texttt{gemma-4-31b-it} \cite{gemma4_google2026}. We note that the generated traces are \emph{post-hoc}: BOCS discovers optimal combinations, not the teacher model. We address this by providing the teacher with a \emph{privileged scorecard}: the top-10 BOCS observations (ranked by mean PAWC) alongside the baseline zero-vector PAWC. This scorecard helps anchor the trace in empirical observations rather than speculation. The teacher is then instructed to produce reasoning grounded entirely in document content---never citing the scores directly---ensuring the student learns to reason from observable document features, not privileged information it will never see at inference time. 

The chosen prompt receives the full scorecard (top-10 combinations plus baseline), while the rejected prompt receives a \emph{truncated} scorecard containing the top-10 combinations at or below $\mathbf{x}^-$'s score. Both prompts additionally receive the full document corpus $D_{\alpha}$ and target document index $j$. Importantly, the teacher does not know $\mathbf{x}^-$ is suboptimal relative to $\mathbf{x}^+$. This produces a rejected trace that is \emph{locally rational} and forces genuine reasoning within an incomplete information set. We include the full teacher prompt in Appendix~\ref{sec:appendix_teacher_prompt}.

\subsubsection{Preference Tuning}
We fine-tune \texttt{gemma-4-E2B-it} \cite{gemma4_google2026} as the student selector with LoRA \citep{hu2022lora} from preference pairs and reasoning traces generated following \S\ref{sec:hard_neg} and \S\ref{sec:cot}. We utilize the Gemma~4 family's native thinking channel for reasoning, and wrap the strategy combination recommendation in \texttt{<final\_answer>} tags for easy parsing. We first perform SFT on chosen traces to ensure format compliance and preference for chosen completions. We then apply DPO \citep{rafailov2023dpo} with a length-normalized loss \citep{meng2024simpo} to maximize the probability difference between chosen and rejected completions. Full set of hyperparameters are listed in Appendix~\ref{sec:appendix_dpo}. 

During training and evaluation, the selector receives: (1)~the query, (2)~all source documents with the target identified, and (3)~the full strategy menu with descriptions. Crucially, the model receives \emph{no} adoption rate or BOCS scores. It must learn to infer competitive pressure directly from document content---detecting signs of prior optimization (eg. added statistics) in competitor sources. This design ensures the selector generalizes to deployment settings where true $\alpha$ is unknown.

\begin{table*}[!t]
\centering
\fontsize{8.5}{9}\selectfont
\setlength{\tabcolsep}{3pt}
\begin{tabular}{lccccccc}
\toprule
& \multicolumn{3}{c}{\textbf{gpt-oss-120b}} & & \multicolumn{3}{c}{\textbf{Llama-3.3-70B-Instruct}} \\
\cmidrule(lr){2-4} \cmidrule(lr){6-8}
\textbf{Method} & \textbf{PAWC} & \textbf{Pos Count} & \textbf{Word Count} & & \textbf{PAWC} & \textbf{Pos Count} & \textbf{Word Count} \\
\midrule
No Rewrite & $17.99 \pm 0.15$ & $17.89 \pm 0.13$ & $18.00 \pm 0.17$ & & $19.16 \pm 0.02$ & $19.12 \pm 0.04$ & $19.06 \pm 0.04$ \\
\midrule
Fluency Optimization & $20.80 \pm 0.15$ & $20.73 \pm 0.14$ & $20.73 \pm 0.15$ & & $20.27 \pm 0.15$ & $20.35 \pm 0.14$ & $20.17 \pm 0.16$ \\
Unique Words & $19.23 \pm 0.09$ & $19.13 \pm 0.09$ & $19.22 \pm 0.14$ & & $18.53 \pm 0.05$ & $18.66 \pm 0.05$ & $18.56 \pm 0.04$ \\
Authoritative & $20.16 \pm 0.06$ & $20.10 \pm 0.04$ & $20.09 \pm 0.07$ & & $19.73 \pm 0.11$ & $19.80 \pm 0.11$ & $19.64 \pm 0.09$ \\
Quotation Addition & $19.60 \pm 0.13$ & $19.54 \pm 0.11$ & $19.50 \pm 0.13$ & & $21.55 \pm 0.08$ & $21.56 \pm 0.06$ & $21.48 \pm 0.07$ \\
Citing Sources & $20.70 \pm 0.13$ & $20.70 \pm 0.16$ & $20.61 \pm 0.13$ & & $20.34 \pm 0.08$ & $20.28 \pm 0.06$ & $20.30 \pm 0.08$ \\
Easy Understanding & $21.09 \pm 0.08$ & $20.93 \pm 0.08$ & $21.07 \pm 0.10$ & & $20.22 \pm 0.12$ & $20.19 \pm 0.11$ & $20.13 \pm 0.11$ \\
Technical Terms & $20.46 \pm 0.10$ & $20.38 \pm 0.10$ & $20.43 \pm 0.10$ & & $18.28 \pm 0.09$ & $18.33 \pm 0.08$ & $18.40 \pm 0.09$ \\
Keyword Stuffing & $19.51 \pm 0.07$ & $19.41 \pm 0.09$ & $19.60 \pm 0.10$ & & $20.68 \pm 0.09$ & $20.67 \pm 0.08$ & $20.70 \pm 0.11$ \\
Statistics Addition & $21.65 \pm 0.17$ & $21.55 \pm 0.18$ & $21.56 \pm 0.14$ & & $21.27 \pm 0.12$ & $21.27 \pm 0.12$ & $21.24 \pm 0.10$ \\
All In One & $21.00 \pm 0.04$ & $20.90 \pm 0.03$ & $20.92 \pm 0.02$ & & $20.75 \pm 0.09$ & $20.89 \pm 0.08$ & $20.65 \pm 0.09$ \\
LLM Guidance & $20.82 \pm 0.08$ & $20.73 \pm 0.06$ & $20.90 \pm 0.09$ & & $21.08 \pm 0.06$ & $21.23 \pm 0.05$ & $21.06 \pm 0.08$ \\
Expanding Facts & $21.24 \pm 0.10$ & $21.14 \pm 0.12$ & $21.06 \pm 0.12$ & & $21.53 \pm 0.07$ & $21.34 \pm 0.07$ & $21.25 \pm 0.09$ \\
Structured Formatting & $21.97 \pm 0.15$ & $21.96 \pm 0.15$ & $21.85 \pm 0.16$ & & $20.00 \pm 0.11$ & $20.10 \pm 0.10$ & $20.07 \pm 0.10$ \\
Balanced View & $18.68 \pm 0.12$ & $18.64 \pm 0.11$ & $18.98 \pm 0.12$ & & $19.67 \pm 0.14$ & $19.61 \pm 0.15$ & $19.78 \pm 0.13$ \\
Conciseness & $21.10 \pm 0.11$ & $21.03 \pm 0.12$ & $21.26 \pm 0.10$ & & $19.89 \pm 0.06$ & $20.17 \pm 0.07$ & $19.92 \pm 0.08$ \\
Pros and Cons & $19.72 \pm 0.09$ & $19.66 \pm 0.09$ & $19.95 \pm 0.07$ & & $21.27 \pm 0.06$ & $21.21 \pm 0.06$ & $21.37 \pm 0.05$ \\
\midrule
AutoGEO & $24.75 \pm 0.08$ & $24.57 \pm 0.09$ & $24.70 \pm 0.06$ & & $21.99 \pm 0.09$ & $22.22 \pm 0.09$ & $21.81 \pm 0.09$ \\
AgenticGEO & $25.20 \pm 0.13$ & $25.10 \pm 0.12$ & $25.06 \pm 0.13$ & & $23.61 \pm 0.05$ & $23.72 \pm 0.06$ & $23.32 \pm 0.06$ \\
\midrule
Competitor-Aware GEO (Ours) & $\mathbf{29.93 \pm 0.09}$ & $\mathbf{30.07 \pm 0.09}$ & $\mathbf{29.79 \pm 0.08}$ & & $\mathbf{25.87 \pm 0.03}$ & $\mathbf{26.18 \pm 0.03}$ & $\mathbf{25.73 \pm 0.05}$ \\
\bottomrule
\end{tabular}
\caption{Overall performance on $\texttt{geo-bench}_{\text{comp}}$ (competitive corpus, averaged across adoption rates $\alpha \in \{0.0, 0.2, 0.4, 0.6, 0.8\}$). We report mean and standard deviation across 1 rewrite (temperature=0) and 5 evaluations from the generative engine (temperature=0.7).}
\label{tab:results_competitive}
\end{table*}

\section{Experimental Setup}
\label{sec:experiments}
We use the \texttt{geo-bench} training split \citep{aggarwal2024geo} to generate $\texttt{geo-bench}_{comp}$ following the competitor simulation described in \S\ref{sec:competitor}.  All document rewrites---both for competitor corpus construction and target document rewriting---are performed using \texttt{gpt-oss-120b} \cite{agarwal2025gpt}. We evaluate performance under two generative engines: \texttt{gpt-oss-120b} and Llama-3.3-70B-Instruct \citep{llama3}.

During evaluation, the fine-tuned selector generates (temperature=0.1) a reasoning trace followed by strategy recommendation, which is parsed from the \texttt{<final\_answer>} tags of the completion. The target document is rewritten once at temperature 0 according to the strategy combination, using the composed prompt in Appendix \ref{sec:appendix_composition}. We generate 5 independent responses from the generative engine at temperature 0.7, reporting mean and standard deviation across three citation-based impression metrics: \textbf{PAWC} (Position-Adjusted Word Count), \textbf{Pos Count} (position-weighted citation frequency), and \textbf{Word Count} (raw cited word count). Formal definitions are given in Appendix~\ref{sec:appendix_metrics}. We compare against all 15 individual rewriting strategies composing our combinatorial search space  (enumerated in \S\ref{sec:data_collection}) as well as existing agentic methods AutoGEO \citep{autogeo} and AgenticGEO \citep{agenticgeo}. We report metrics on both \texttt{geo-bench} and $\texttt{geo-bench}_{comp}$. We benchmark LLM-based impression metrics from \citet{mageo} as well as transferability to E-Commerce and Researchy-GEO datasets \cite{autogeo}.

\section{Results}
\label{sec:results}

\paragraph{Main Results.}
Table~\ref{tab:results_original} reports mean PAWC, position count, and word count on the original \texttt{geo-bench} corpus and Table~\ref{tab:results_competitive} on $\texttt{geo-bench}_{\text{comp}}$, each evaluated under two generative engines: \texttt{gpt-oss-120b} and Llama-3.3-70B-Instruct. Competitor-Aware GEO outperforms all baselines across both datasets and evaluation models. On the original corpus (Table~\ref{tab:results_original}), our method achieves $32.62$ PAWC on \texttt{gpt-oss-120b} and $29.55$ on Llama~3.3 70B---a $+4.67$ and $+2.75$ PAWC improvement over the strongest baseline AgenticGEO. These gains hold across all three impression metrics. Notably, our result ($32.62$) recovers $84\%$ of the Oracle BOCS ceiling ($39.00$) -- a theoretical maximum obtained from independently running BOCS on each test data point. We also ablate both reasoning traces in the training dataset and top-10 BOCS scorecard inclusion in teacher reasoning generation, demonstrating that training on \textit{grounded} reasoning traces is crucial in generalizing to unseen document corpora. We include the full results and analysis in Appendix \ref{sec:appendix_ablatereasoning} and Table~\ref{tab:ablation_reasoning}. 

On $\texttt{geo-bench}_{\text{comp}}$, the no-rewrite baseline drops to $17.99$, indicating that unoptimized content is actively penalized when competitors rewrite. Despite this compression, our method achieves $29.93$ PAWC, outperforming agentic methods by $>18\%$. The gap widens under competition: the margin over AgenticGEO grows $2.1\%$ between \texttt{geo-bench} and $\texttt{geo-bench}_{\text{comp}}$, demonstrating that competitor-awareness becomes increasingly valuable as optimization saturates the corpus.

\begin{table*}[h]
\centering
\footnotesize
\setlength{\tabcolsep}{4pt}
\begin{tabular}{lcccc}
\toprule
\textbf{Method} & \textbf{Length Ratio} & \textbf{Attribution Accuracy} & \textbf{Faithfulness} & \textbf{Key Point Coverage} \\
\midrule
No Rewrite          & $ -$ & $\mathbf{6.15}$ & $\mathbf{8.29}$ & $6.60$ \\
\midrule
Quotation Addition  & $1.21$ & $3.92$ & $5.08$ & $6.81$ \\
LLM Guidance        & $1.78$ & $5.47$ & $7.90$ & $7.30$ \\
Conciseness         & $0.49$ & $5.80$ & $7.99$ & $6.89$ \\
\midrule
AutoGEO             & $2.21$ & $5.52$ & $6.58$ & $7.47$ \\
AgenticGEO          & $\textbf{0.91}$ & $5.15$ & $6.71$ & $7.15$ \\
\midrule
Competitor-Aware GEO (Ours) & $1.29$ & $4.18$ & $4.90$ & $\mathbf{7.48}$ \\
\bottomrule
\end{tabular}
\caption{LLM-based qualitative metrics on \texttt{geo-bench} (original corpus, \texttt{gpt-oss-120b}). Metrics from the MAGEO evaluation framework \cite{mageo}: Attribution Accuracy, Faithfulness, and Key Point Coverage (higher is better). Length Ratio is relative to the no-rewrite baseline; bold indicates best per column. We report mean across 1 rewrite (temperature=0) and 5 independent evaluations (temperature=0.7) from the generative engine.}
\label{tab:results_llm_metrics}
\end{table*}

\begin{figure}[t]
    \centering
    \includegraphics[width=\columnwidth]{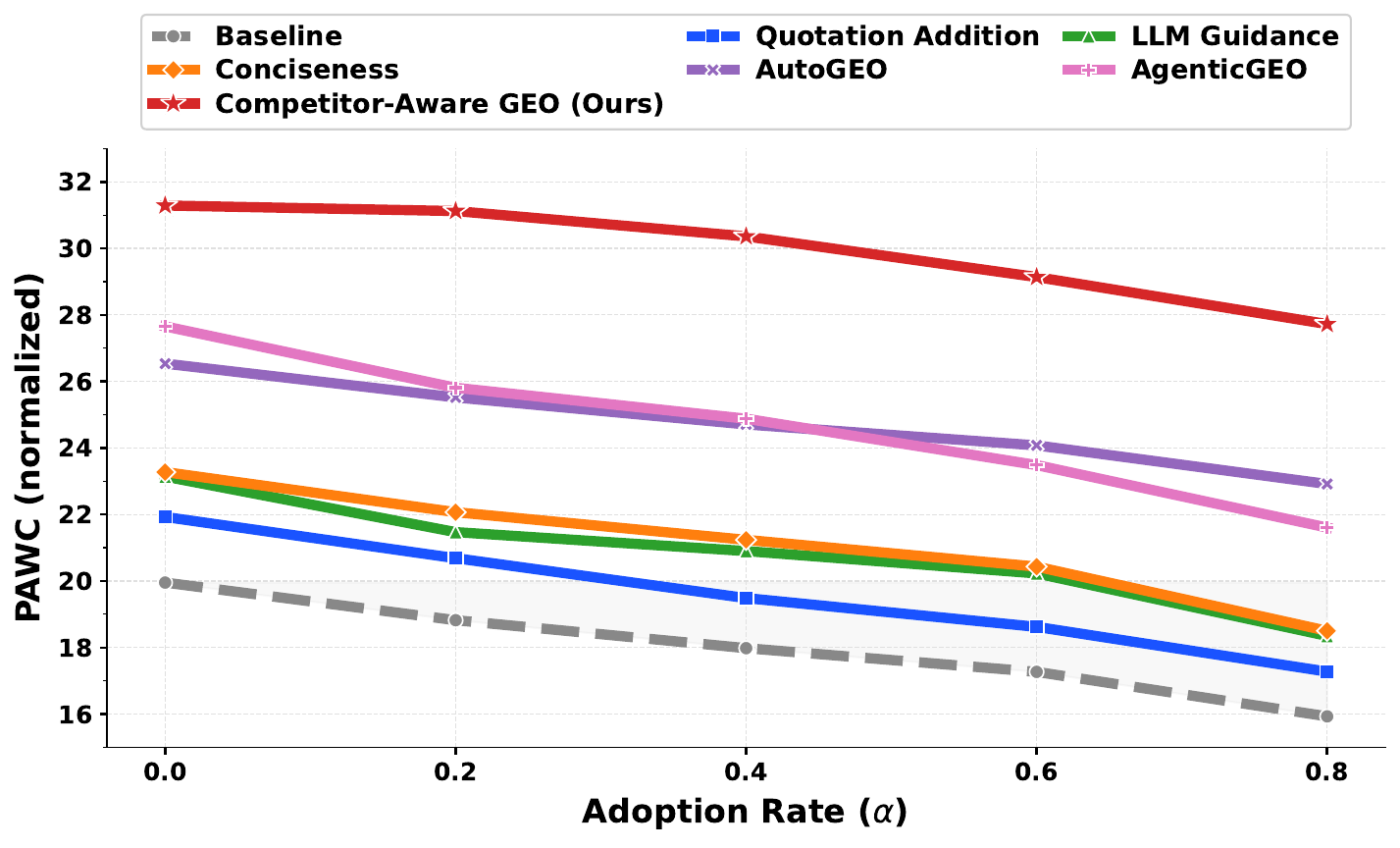}
    \caption{Mean PAWC over various adoption rates $\alpha \in \{0,0.2,0.4,0.6,0.8\}$ on $\texttt{geo-bench}_{comp}$. We evaluate our Competitor-Aware GEO method against static single-strategy and agentic baselines. We rewrite once (temperature=0) and evaluate 5 times (temperature=0.7), both with gpt-oss-120b.}
    \label{fig:pawc_vs_alpha}
    \vspace{-1em}
\end{figure}
% \vspace{-2em}

\paragraph{Performance Scales with Adoption Rate.}
To characterize how our selector adapts to increasing competitor optimization, we plot mean PAWC as a function of adoption rate $\alpha \in \{0.0, 0.2, 0.4, 0.6, 0.8\}$ (Figure~\ref{fig:pawc_vs_alpha}). All baselines exhibit monotonically decreasing PAWC as $\alpha$ grows, and single-strategy methods at full competitive optimization ($\alpha=0.8$) drop below the unoptimized ($\alpha=0$) baseline. In contrast, Competitor-Aware GEO exhibits the slowest degradation of any method, declining $11.4\%$ across the full adoption range. Notably, we note no statistically significant  performance drop from $\alpha{=}0.0$ to $\alpha{=}0.2$. By inferring alpha from competitive context at inference time, our selector shifts strategy combinations toward less saturated alternatives as the competitive landscape is increasingly optimized.

\begin{table}[h]
\centering
\footnotesize
\setlength{\tabcolsep}{4pt}
\begin{tabular}{lcc}
\toprule
\textbf{Method} & \textbf{E-Comm.} & \textbf{Researchy-GEO} \\
\midrule
No Rewrite          & $17.94 \pm 0.35$ & $19.17 \pm 0.16$ \\
\midrule
Quotation Addition      & $19.46 \pm 0.25$ & $23.09 \pm 0.19$ \\
LLM Guidance            & $23.17 \pm 0.58$ & $25.66 \pm 0.15$ \\
Conciseness             & $23.53 \pm 0.15$ & $25.32 \pm 0.23$ \\
\midrule
AutoGEO                 & $26.37 \pm 0.77$ & $30.98 \pm 0.71$ \\
AgenticGEO              & $34.96 \pm 0.53$ & $27.54 \pm 0.33$ \\
\midrule
Ours & $\mathbf{37.78 \pm 1.10}$ & $\mathbf{33.48 \pm 0.24}$ \\
\bottomrule
\end{tabular}
\caption{Overall transferability to out-of-distribution datasets. Our model is trained on \texttt{geo-bench} only and applied zero-shot. We report mean and standard deviation of PAWC across 1 rewrite and 5 independent evaluations with \texttt{gpt-oss-120b}.}
\label{tab:results_cseobench}
\vspace{-1em}

\end{table}

\paragraph{Competitor-Aware GEO Transfers to Out-of-Distribution Datasets}
Table~\ref{tab:results_cseobench} reports zero-shot transfer results on two datasets with distinct query and document distributions: E-Commerce and Researchy-GEO. Our method achieves the highest PAWC on both datasets, improving upon the strongest existing method by $>8\%$. We note that transferability is stronger among content-conditioned methods, allowing strategy selection to adapt to the characteristics of each domain without retraining. Our competitor-aware method further improves cross-domain generalization as it reasons over the full document context, exploring dynamics and exploiting structural or content weaknesses. We hypothesize that such capabilities transfer more robustly than overarching heuristics or conditioning on target document content alone, since document dynamics are a property of the retrieval context rather than any particular query distribution.

\paragraph{PAWC Optimization Introduces a Faithfulness--Enrichment Trade-off.}
Table~\ref{tab:results_llm_metrics} reports LLM-based impression metrics from the MAGEO evaluation framework \cite{mageo} (detailed in Appendix \ref{sec:appendix_metrics}) on the \texttt{geo-bench} corpus, evaluated using \texttt{gpt-oss-120b}. Our method achieves the highest Key Point Coverage (7.48), but comes at the cost of Faithfulness (4.90) and Attribution Accuracy (4.18). We argue that this is an expected and intentional trade-off inherent to generative rewriting. Since both metrics measure divergence from the original source rather than factual correctness, they cannot separate hallucination from verifiable enrichment. Thus, they form a lower bound on true faithfulness. This dynamic is illustrated by the Conciseness baseline, which achieves the highest Faithfulness score, but shortens texts by $>50\%$ and performs worse than all agentic methods on Key Point Coverage. We note this is a limitation in current automated evaluations, and future work must distinguish between hallucinated and augmented content.

\section{Conclusion}
\label{sec:conclusion}
In this work, we introduce the first method to formalize and evaluate GEO as a competitor-aware combinatorial strategy selection problem. Our two-phase pipeline---BOCS-driven search over the rewriting strategy space followed by preference tuning with score-grounded reasoning traces---produces a selector that analyzes the competitive document landscape and recommends tailored rewriting strategies without access to adoption rate labels or privileged information at inference time. Our method significantly outperforms existing single-strategy and agentic baselines on \texttt{geo-bench} and $\texttt{geo-bench}_{\text{comp}}$, exhibiting the slowest performance degradation across adoption rates as the selector learns to infer competitive pressure from the full document corpus. Robust transferability to other datasets confirms that reasoning over document dynamics generalizes across domains. Broadly, this work points toward a new paradigm for content optimization in generative settings---one where strategy selection is inherently multi-agent, corpus-conditioned, and adaptive.

\section*{Limitations}

Our evaluation relies on LLM-generated responses as a proxy for real generative engine outputs, which may not fully reflect production system behavior. Similarly, our competitor simulation uses synthetic distributions rather than observed real-world adoption patterns. Beyond these evaluation proxies, our 15-strategy space is not exhaustive, despite covering the strategies identified in prior GEO work. Furthermore, we primarily evaluate rewrites in a controlled generation context rather than comprehensively simulating the full retrieval and re-ranking pipeline. This limitation, coupled with an observed drop in faithfulness compared to existing baselines, points to an important direction for future research, which would incorporate both retrievability and faithfulness as explicit constraints within the optimization formulation. Finally, our method operates with an information advantage over prior GEO approaches by ingesting the full competitive corpus at inference time, whereas previous approaches unconditionally apply rewriting strategies or only condition on the target content. We note, however, that this information is directly observable in practice—a content producer can readily inspect competing documents on a search result page—and conditioning on it is precisely what enables competitor-aware strategy selection.

\section*{Ethics Statement}
Our method is designed to help content creators adapt to generative search by improving the visibility of high-quality, relevant material. The same techniques, however, could be misused to game generative engines in ways that degrade information quality---for instance, by fabricating citations or inserting unsupported claims. To guard against this, our rewriting prompts explicitly forbid inventing quotations, fabricating sources, or introducing unverifiable numbers, and we regard faithfulness-constrained optimization as a key direction for future work. We intend this research to surface trustworthy content, not to manufacture false signals of credibility. We are committed to the responsible development and application of GEO methods, and future work will continue to investigate and address these ethical concerns.

\section*{Acknowledgments}
We acknowledge the use of Generative AI tools to aid in the writing of this manuscript, conducting literature review, and AI-assisted coding. However, all ideas, final implementations, and interpretations of data originate and are verified thoroughly by the authors of this paper.

\bibliography{custom}

\appendix

\section{Additional Results}

\paragraph{Reasoning with BOCS Scorecard is Necessary for Grounded CoT traces and Effective DPO.}
\label{sec:appendix_ablatereasoning}
Table~\ref{tab:ablation_reasoning} isolates the contribution of reasoning traces and preference optimization by comparing training variants on \texttt{geo-bench} with \texttt{gpt-oss-120b} as the generative engine. Without reasoning traces, DPO provides no benefit over SFT. We attribute this to the supervision signal when preference pairs consist solely of strategy label lists and do not meaningfully differ. Including reasoning traces without a BOCS scorecard improves performance from no-reasoning baselines. However, we note the largest improvement performing SFT and subsequently DPO on reasoning traces generated with grounded results from a BOCS scorecard. The model learns to condition its strategy recommendations on observable document features rather than memorizing label frequencies. Adding DPO on top of SFT with reasoning yields a further $+3.35$ PAWC to $32.62 \pm 0.25$, confirming that the contrastive signal becomes effective only when chosen and rejected responses differ in a rich reasoning trace.

\begin{table}[h]
\centering
\fontsize{9}{9}\selectfont
\begin{tabular}{lc}
\toprule
\textbf{Method} & \textbf{PAWC} \\
\midrule
Zero-shot teacher  & $21.53$ \\
SFT w/o reasoning      & $24.12$ \\
SFT+DPO, w/o reasoning       & $23.92$ \\
SFT + DPO, w reasoning   & $26.96$ \\
SFT w scorecard-grounded reasoning    & $29.27$ \\
SFT+DPO, w scorecard-grounded reasoning   & $\mathbf{32.62}$ \\
\midrule
\color{gray}\textit{Oracle BOCS (upper bound)$^\dagger$}   & \color{gray}\textit{39.00} \\
\bottomrule
\end{tabular}
\caption{Ablation on reasoning traces. All variants trained on \texttt{geo-bench} with rewriting and evaluation performed with \texttt{gpt-oss-120b}. $^\dagger$Oracle BOCS runs BOCS independently per test data point---computationally intractable at deployment, shown as theoretical ceiling only.}
\label{tab:ablation_reasoning}
\end{table}

\paragraph{BOCS Order Ablation} We select a second-order BOCS surrogate to capture pairwise strategy interactions, but validate this choice empirically by comparing against a linear (order-1) surrogate. Table~\ref{tab:ablation_bocs_order} reports mean best PAWC discovered per data point across the full \texttt{geo-bench} training split ($n{=}7{,}998$). The order-2 surrogate achieves a mean PAWC of $31.02$ versus $30.63$ for order-1, a $+0.39$ improvement. This gap confirms that pairwise interaction terms---which model synergies and conflicts between strategies (e.g., Conciseness suppressing the benefit of Expanding Facts)---provide signal that a linear surrogate cannot capture, enabling BOCS to converge to higher-quality combinations within the same evaluation budget. Full hyperparameters are found in Table \ref{tab:bocs_params}.

\begin{table}[h]
\centering
\small
\begin{tabular}{lcc}
\toprule
\textbf{Surrogate} & \textbf{Features} & \textbf{PAWC} \\
\midrule
Order-1 (linear)   & 16  & $30.63$ \\
Order-2 (pairwise) & 121 & $\mathbf{31.02}$ \\
\bottomrule
\end{tabular}
\caption{Order-1 vs.\ order-2 BOCS surrogate on \texttt{geo-bench} training split ($n{=}7{,}998$). Mean best PAWC found per data point averaged across the full training split.}
\label{tab:ablation_bocs_order}
\end{table}

\paragraph{Teacher Reasoning Trace Audit}
\label{sec:appendix_leakage}
Although the teacher is instructed to ground its reasoning entirely in document content and never cite the BOCS scorecard directly (\S\ref{sec:cot}), this step relies on the instruction-following capabilities of the teacher model. To quantify residual leakage, we audit all chosen and rejected reasoning traces in the \texttt{geo-bench} DPO training set. We feed the full teacher prompt, the top-10 BOCS scorecard, and the generated reasoning trace to \texttt{gpt-oss-120b} and ask it to flag any trace that references the scorecard, its rankings, or numerical scores. Results over all 12{,}124 traces (6{,}062 chosen and 6{,}062 rejected) are reported in Table~\ref{tab:leakage_audit}. Even with explicit instructions not to leak, the teacher references rankings or the scorecard in $\approx$1\% of traces. Crucially, the student never observes the scorecard at inference time, and the strong downstream DPO performance (Table~\ref{tab:ablation_reasoning}) indicates the student generalizes well despite training on a small fraction of traces that may carry leaked artifacts. Nevertheless, future work using teacher-distilled reasoning for preference learning should apply a post-hoc audit and filter leaked traces before training.

\begin{table}[h]
\centering
\small
\setlength{\tabcolsep}{4pt}
\begin{tabular}{lrrrrr}
\toprule
\textbf{Split} & \textbf{Traces} & \textbf{Leaked} & \textbf{Rate} & \textbf{Rank.} & \textbf{Score.} \\
\midrule
Chosen  & 6{,}062 & 61 & 1.00\% & 22 & 39 \\
Rejected & 6{,}062 & 52 & 0.86\% & 14 & 38 \\
\midrule
\textbf{Total}    & \textbf{12{,}124} & \textbf{113} & \textbf{0.93\%} & \textbf{36} & \textbf{77} \\
\bottomrule
\end{tabular}
\caption{Leakage audit of teacher-generated reasoning traces on the \texttt{geo-bench} DPO training set, judged by \texttt{gpt-oss-120b}. \textbf{Leaked} counts traces referencing privileged scorecard information; \textbf{Rank.} and \textbf{Score.} count traces referencing scorecard rankings and raw scores, respectively (a trace may reference both).}
\label{tab:leakage_audit}
\end{table}

\section{Evaluation Metrics}
\label{sec:appendix_metrics}

\subsection{Citation-Based Metrics}
Let $R$ consist of sentences $\{s_1, \ldots, s_N\}$, and let $\text{cites}(s_i)$ denote the multiset of 1-based citation indices extracted from sentence $s_i$. Define $c_j(s_i) = \sum_{c \in \text{cites}(s_i)} \mathds{1}[c = j]$ as the number of citations to source $j$ in sentence $s_i$, $|s_i|$ as the word count of sentence $s_i$ (words with more than 2 characters), and $w_i = \exp(-i/(N{-}1))$ as the exponential position decay weight.

\paragraph{PAWC (Position-Adjusted Word Count).}
The primary optimization objective, weighting cited word count by sentence position:
\begin{equation}
    \text{PAWC}(j, R) = \sum_{i=1}^{N} \frac{c_j(s_i) \cdot |s_i| \cdot w_i}{|\text{cites}(s_i)|}
\end{equation}
Earlier sentences receive higher weight via the exponential decay $w_i$, reflecting that citations near the beginning of a response carry greater visibility.

\paragraph{Pos Count (Position-Weighted Citation Frequency).}
Measures how often the source is cited, weighted by position, without scaling by word count:
\begin{equation}
    \text{PosCount}(j, R) = \sum_{i=1}^{N} \frac{c_j(s_i) \cdot w_i}{|\text{cites}(s_i)|}
\end{equation}

\paragraph{Word Count (Raw Cited Word Count).}
Counts the total words from the source that appear in cited sentences, without position adjustment:
\begin{equation}
    \text{WordCount}(j, R) = \sum_{i=1}^{N} \frac{c_j(s_i) \cdot |s_i|}{|\text{cites}(s_i)|}
\end{equation}
Together, PAWC combines the citation frequency and lexical richness signals from Pos Count and Word Count while additionally rewarding early-position citations.

\subsection {Length-Based Metrics}

\paragraph{Length Ratio (LR).}
  Measures the relative change in document length introduced by a rewriting method, defined as the ratio of the rewritten document's word count to that of the original source document:
  \begin{equation}                            
      \text{LR}(d', d) = \frac{|d'|_w}{|d|_w}
  \end{equation}
  where $|\cdot|_w$ denotes word count, $d$ is the original document, and $d'$ is its rewritten document. The average length ratio across a benchmark is reported as:                                        
  \begin{equation}
      \overline{\text{LR}} = \frac{1}{M} \sum_{m=1}^{M} \frac{|d'_m|_w}{|d_m|_w}                                                                                                                                
  \end{equation}                                                                       
  A value of $\overline{\text{LR}} = 1$ indicates length-neutral rewriting, values above 1 reflect expansion, and values below 1 reflect compression. We report $\overline{\text{LR}}$ as a diagnostic metric
  alongside quality scores, since methods that substantially expand document length may inflate citation-based and key-point coverage metrics without genuine quality improvement.

 \subsection{Qualitative Metrics via LLM Evaluation}

  In addition to the position-based citation metrics above, we adopt three qualitative dimensions from the MAGEO evaluation framework~\cite{mageo} to assess the semantic and epistemic quality of        
  rewritten content. These metrics are scored on a 1--10 scale by an LLM judge \texttt{gpt-oss-120b} , using the prompt template in Appendix~\ref{box:mageo_evaluator_prompt}.

  \paragraph{AA (Attribution Accuracy):}                                               
  Measures whether factual claims in the rewritten content are traceable to verifiable sources or citations. A candidate receives a high AA score when its assertions are explicitly grounded in attributed evidence;
   it is penalized when claims are unsupported or attribution is absent.

  \paragraph{FA (Response-level Faithfulness):}
  Measures semantic preservation between the rewritten candidate and the original source document treated as a safety-critical dimension alongside AA. A candidate is penalized when it introduces hallucinated facts, omits critical information, or distorts the
  meaning of the original.

  \paragraph{KC (Key-Point Coverage):}
  Measures the extent to which the rewritten content addresses the user query with sufficient factual depth and topical completeness. A candidate scores high on KC when it directly answers the posed question with
  relevant, specific information; it scores low when the response is generic or omits key query-relevant facts.

\begin{promptbox}{Prompt Template for Qualitative Evaluation}
\label{box:mageo_evaluator_prompt}
\textbf{[SYSTEM PROMPT]} \\
You are the Evaluator Agent in MAGEO. Judge each candidate under the DSV-CF framework and return JSON only. \\
\\
Score each of the 8 metrics on a 1--10 scale using these exact definitions: \\
\\
\textbf{SSV metrics (Surface \& Structural Value):} \\
- \textbf{wlv} — Word-Level Value: fluency, grammar, and readability of the rewritten text. \\
- \textbf{dpa} — Document-Position Awareness: how well the content is structured and positioned for visibility in generative engine responses. \\
- \textbf{cp} — Content Preservation: how well the original meaning, topic, and facts are retained without distortion. \\
- \textbf{si} — Semantic Integrity: coherence and logical consistency of the rewritten content. \\
\\
\textbf{ISI metrics (Information-Source Integrity):} \\
- \textbf{aa} — Attribution Accuracy: an LLM judge extracts claims attributed to the target document and checks whether they are entailed by the original source. This is the main anti-hallucination safeguard. Score low if claims are unsupported or fabricated. \\
- \textbf{fa} — Response-level Faithfulness: checks whether the optimized document remains semantically faithful to the original document and has not introduced unsupported content during editing. Score low if new facts not in the source were introduced. \\
- \textbf{kc} — Key-Point Coverage: the system first extracts key information points from the target document; this metric calculates the recall of those key points within the response, measuring how much substance was successfully transferred. Score low if important points are missing. \\
- \textbf{ad} — Answer Dominance: designed for comparative or recommendation queries, this metric determines if the target document is presented as the primary solution. The LLM analyzes the sentiment and recommendation strength of the response relative to competing sources. \\
\\
Penalize candidates when attribution is weak, faithfulness is poor, or key points are missing. \\
\\
\textbf{[USER PROMPT]} \\
\textbf{User query:} \\
\texttt{\{user\_query\}} \\
\\
\textbf{Baseline content:} \\
\texttt{\{baseline\_content\}} \\
\\
\textbf{Candidate list:} \\
\texttt{\{candidates\}} \\
\\
\textbf{Engine preference profile:} \\
\texttt{\{engine\_rules\}} \\
\\
For each candidate, predict DSV-CF-related scores and short comments. \\
Return JSON only with this schema: \\
\texttt{\{} \\
\texttt{~~"evaluations": [} \\
\texttt{~~~~\{} \\
\texttt{~~~~~~"candidate\_id": "V1",} \\
\texttt{~~~~~~"predicted\_scores": \{ "wlv": 6.5, "dpa": 6.2, "cp": 7.0, "si": 6.8, "aa": 8.5, "fa": 8.0, "kc": 7.2, "ad": 6.9 \},} \\
\texttt{~~~~~~"metric\_critic\_comment": "...",} \\
\texttt{~~~~~~"safety\_critic\_comment": "...",} \\
\texttt{~~~~~~"preference\_critic\_comment": "...",} \\
\texttt{~~~~~~"overall\_comment": "..."} \\
\texttt{~~~~\}} \\
\texttt{~~]} \\
\texttt{\}}
\end{promptbox}

\begin{promptbox}{Example Engine Preference Profile (\texttt{\{engine\_rules\}} Injection)}
\label{box:engine_rules_profile}
\texttt{\{} \\
\texttt{~~"engine\_id": "gpt-oss-120b",} \\
\texttt{~~"format\_preferences": [} \\
\texttt{~~~~"Use a clear hierarchical heading structure (H1-H3) with query keywords; keep hierarchy shallow",} \\
\texttt{~~~~"Present enumerations, comparisons, and quantitative data in bullet lists, numbered lists, or tables",} \\
\texttt{~~~~"Place a concise self-contained answer (<=150 words) at the very beginning followed by a Key Takeaway summary",} \\
\texttt{~~~~"Keep overall document within <=500 words or <=800 tokens; core answer passage within <=200 tokens",} \\
\texttt{~~~~"Write each sentence as a short (<=30 words) declarative statement conveying a single fact"} \\
\texttt{~~],} \\
\texttt{~~"content\_preferences": [} \\
\texttt{~~~~"Answer every explicit or implicit sub-question; enumerate requested items without omission or duplication",} \\
\texttt{~~~~"Ensure at least 80\% of tokens directly address the query; remove filler, promotional, or navigation text",} \\
\texttt{~~~~"Expand acronyms and domain-specific terms at first occurrence; use consistent terminology throughout",} \\
\texttt{~~~~"Maintain neutral, factual tone; exclude promotional, speculative, or opinionated language",} \\
\texttt{~~~~"Write in plain English at approximately grade 8 reading level with correct grammar and spelling"} \\
\texttt{~~],} \\
\texttt{~~"citation\_preferences": [} \\
\texttt{~~~~"Achieve citation density of at least 80\% of factual statements",} \\
\texttt{~~~~"Provide inline machine-readable citation for every factual or quantitative claim including author, title, date, and URL",} \\
\texttt{~~~~"Display author name, credentials, source outlet, and publication date; ensure sources are recent (<=5 years)",} \\
\texttt{~~~~"Verify all hyperlinks are functional and point to reputable publicly accessible sources"} \\
\texttt{~~],} \\
\texttt{~~"risk\_constraints": [} \\
\texttt{~~~~"Exclude promotional, speculative, or opinionated language unless explicitly requested",} \\
\texttt{~~~~"Isolate non-answer elements (ads, navigation links, UI labels, emojis) or remove from core answer block",} \\
\texttt{~~~~"Ensure content complies with safety and policy guidelines",} \\
\texttt{~~~~"Include SEO-friendly metadata mirroring query terms in title, headings, and early sentences"} \\
\texttt{~~]} \\
\texttt{\}}
\end{promptbox}

\section{BOCS Hyperparameters}
\label{sec:appendix_bocs}
We include the full hyperparameter details for running BOCS in Table \ref{tab:bocs_params}.

\begin{table}[h]
  \centering
  \small
  \setlength{\tabcolsep}{4pt}
  \begin{tabular}{lc}
    \toprule
    \textbf{Parameter} & \textbf{Value} \\
    \midrule
    Surrogate order & 2 \\
    Evaluation budget & 50 \\
    Initial random points & 10 \\
    Early stopping patience & 10 \\
    SA iterations & 200 \\
    SA cooling rate & 0.85 \\
    SA reruns & 5 \\
    Gibbs samples & 1000 \\
    \bottomrule
  \end{tabular}
  \caption{BOCS hyperparameters.}
  \label{tab:bocs_params}
\end{table}

\section{Fine-Tuning Training Details}
\label{sec:appendix_dpo}
We fine-tune \texttt{gemma-4-E2B-it} using the TRL library \citep{vonwerra2020trl} for both the SFT and DPO stages. SFT is performed on chosen reasoning traces to establish format compliance, followed by DPO with a length-normalized loss to learn from contrastive preference pairs. Both stages use LoRA \citep{hu2022lora} for fine-tuning. We merge the LoRA adapters into the base model after SFT and initialize new adapters for the DPO phase. During DPO, we enable gradient checkpointing to reduce peak memory usage. Full hyperparameters are listed in Table~\ref{tab:dpo_params}.

\begin{table}[h]
  \centering\small
  \begin{tabular}{lcc}
    \toprule
    \textbf{Parameter} & \textbf{SFT} & \textbf{DPO} \\
    \midrule
    Base model & \multicolumn{2}{c}{\texttt{gemma-4-E2B-it}} \\
    Teacher model & \multicolumn{2}{c}{\texttt{gemma-4-31B-it}} \\
    \midrule
    LoRA rank ($r$) & 16 & 16 \\
    LoRA alpha ($\alpha$) & 16 & 32 \\
    LoRA dropout & 0.0 & 0.05 \\
    DPO $\beta$ & -- & 0.1 \\
    Learning rate & $2{\times}10^{-5}$ & $5{\times}10^{-7}$ \\
    Epochs & 2 & 3 \\
    Batch size & 16 & 16 \\
    Warmup ratio & 0.1 & 0.1 \\
    Grad.\ checkpointing & off & on \\
    Precision & \multicolumn{2}{c}{bfloat16} \\
    \bottomrule
  \end{tabular}
  \caption{SFT and DPO fine-tuning hyperparameters.}
  \label{tab:dpo_params}
\end{table}

\section{Rewriting Strategy Prompts}
\label{sec:appendix_prompts}

Each of the 15 rewriting strategies is implemented as a standalone prompt template. All prompts share a common system prompt establishing the GEO research context, followed by strategy-specific instructions. The \texttt{\{text\}} placeholder is filled with the source document at inference time.

\subsection{System Prompt}

\begin{promptbox}{}
You are an expert ml researcher having previous background in SEO and search engines in general. You are working on novel research ideas for next generation of products. These products will have language models augmented with search engines, with the task of answering questions based on sources backed by the search engine. This new set of systems will be collectively called language engines (generative search engines). This will require websites to update their SEO techniques to rank higher in the llm generated answer. Specifically they will use GEO (Generative Engine Optimization) techniques to boost their visibility in the final text answer outputted by the Language Engine.
\end{promptbox}

\subsection{Individual Strategy Prompts}

\begin{promptbox}{Fluency Optimization}
Task: Rewrite the source to improve fluency and coherence.\\
Constraints:\\
- Do not alter the core content.\\
- Improve sentence transitions and readability.\\
- Keep the structure and length roughly the same.\\
Source: \{text\}\\
Output: The rewritten source text only.
\end{promptbox}

\begin{promptbox}{Unique Words}
Task: Revise the source by using more unique and precise vocabulary.\\
Constraints:\\
- Preserve the original meaning and all core information.\\
- Do not add new claims or remove any content.\\
- Keep the length and structure roughly the same.\\
Source: \{text\}\\
Output: The revised source text only.
\end{promptbox}

\begin{promptbox}{Authoritative}
Task: Make the source sound confident, authoritative, and expert.\\
Constraints:\\
- Do not add new facts or remove any information.\\
- Keep the original structure (formatting, bullets, spacing).\\
- Strengthen tone via wording choices, not by exaggerating or making unverifiable claims.\\
Source: \{text\}\\
Output: The revised source text only.
\end{promptbox}

\begin{promptbox}{Quotation Addition}
Task: Increase perceived authority by adding a few short, relevant quotations from reputable entities (e.g., well-known organizations or experts).\\
Constraints:\\
- Quotes must be accurate and attributable; do not invent quotes.\\
- Do not change core content; keep structure and length similar.\\
- Integrate quotes inline without adding long new paragraphs.\\
Source: \{text\}\\
Output: The revised source text only.
\end{promptbox}

\begin{promptbox}{Cite Sources}
Task: Strengthen credibility by adding a small number of natural-language citations to credible sources (e.g., industry reports, standards, official docs).\\
Constraints:\\
- Citations must be plausible and verifiable; do not fabricate sources.\\
- Do not change the core information or add new claims.\\
- Keep structure and length roughly the same (about 5-6 citations total).\\
Source: \{text\}\\
Output: The revised source text only.
\end{promptbox}

\begin{promptbox}{Easy-To-Understand}
Task: Rewrite the source in simple, easy-to-understand language.\\
Constraints:\\
- Do not omit, add, or alter any core information.\\
- Keep the original structure and roughly the same length.\\
- Only rephrase sentences for clarity and readability.\\
Source: \{text\}\\
Output: The simplified source text only.
\end{promptbox}

\begin{promptbox}{Technical Words}
Task: Rewrite the source in a more technical style using domain-appropriate terminology.\\
Constraints:\\
- Preserve all core information; do not introduce new claims.\\
- Keep the structure and length roughly unchanged.\\
- Rephrase sentences to sound more technical and precise.\\
Source: \{text\}\\
Output: The revised source text only.
\end{promptbox}

\begin{promptbox}{Keyword Stuffing}
Task: Improve the source by inserting up to 10 NEW, relevant SEO keywords that are NOT already present in the text.\\
Constraints:\\
- Do not change, add, or remove any core information.\\
- Keep the original structure (paragraphing, bullet points, line breaks).\\
- Insert keywords naturally inline (no keyword list at the end).\\
Source: \{text\}\\
Output: The updated source text only.
\end{promptbox}

\begin{promptbox}{Statistics Addition}
Task: Add a few concise, relevant statistics or numerical facts to improve concreteness.\\
Constraints:\\
- Statistics must be verifiable; do not invent numbers.\\
- Do not modify core content beyond inserting stats inline.\\
- Keep the original structure and stop at the end of the original source.\\
Source: \{text\}\\
Output: The revised source text only.
\end{promptbox}

\begin{promptbox}{Expanding Facts}
Task: Enrich the source by expanding on key claims with brief explanations of how and why, and by clarifying underlying causes.\\
Constraints:\\
- Do not remove any existing information or change the core meaning.\\
- Keep the original structure (paragraphing, bullets, spacing).\\
- Expansions should be concise and inline; do not add long new sections.\\
Source: \{text\}\\
Output: The revised source text only.
\end{promptbox}

\begin{promptbox}{Structured Formatting}
Task: Reorganize the source using structured formatting (tables, bullet points, numbered lists) where it improves clarity.\\
Constraints:\\
- Preserve all core information; do not add or remove claims.\\
- Only restructure; do not rewrite the language itself beyond minimal adjustments for the new format.\\
- Use formatting that fits the content (e.g., tables for comparisons, bullets for lists).\\
Source: \{text\}\\
Output: The restructured source text only.
\end{promptbox}

\begin{promptbox}{Balanced View}
Task: Revise the source to present a more balanced perspective by acknowledging multiple viewpoints where relevant.\\
Constraints:\\
- Do not remove any existing information or claims.\\
- Keep the structure and length roughly the same.\\
- Add brief counterpoints or alternative perspectives inline without overriding the original stance.\\
Source: \{text\}\\
Output: The revised source text only.
\end{promptbox}

\begin{promptbox}{Conciseness}
Task: Tighten the source by eliminating filler content, verbose language, and repetition.\\
Constraints:\\
- Preserve all core information and key claims.\\
- Do not add new content or change the meaning.\\
- Keep the original structure; only cut unnecessary words and redundant phrases.\\
Source: \{text\}\\
Output: The concise source text only.
\end{promptbox}

\begin{promptbox}{Pros and Cons}
Task: Enhance the source by adding clear pros-and-cons reasoning and comparative analysis to support its recommendations or claims.\\
Constraints:\\
- Do not remove any existing information.\\
- Keep the original structure roughly the same.\\
- Integrate pros/cons and comparative points inline or as brief additions; do not add lengthy new sections.\\
Source: \{text\}\\
Output: The revised source text only.
\end{promptbox}

\begin{promptbox}{LLM Guidance}
Task: Create an LLM-friendly summary of the source and prepend it to the original text. The final output must contain BOTH the summary AND the full original source text, concatenated together.\\
Constraints:\\
- The summary should be a concise markdown overview (title, introduction, key sections) that helps an LLM quickly understand the content.\\
- Do not alter the original source text in any way.\\
- The output format must be: summary first, then a blank line, then the complete original source text verbatim.\\
Source: \{text\}\\
Output: The summary followed by the complete original source text.
\end{promptbox}

\subsection{Multi-Strategy Composition}
\label{sec:appendix_composition}

When BOCS selects a combination $\mathbf{x} \in \{0,1\}^{15}$ with multiple active strategies, we compose a single unified rewriting prompt by extracting the instruction fragment for each active strategy and numbering them as simultaneous objectives. The composition template is:

\begin{promptbox}{Multi-Strategy Composition Template}
Rewrite the following source text by applying ALL of the optimization objectives listed below simultaneously.\\
\\
\textbf{Query Context}\\
The source text is being optimized to better address: \{query\}\\
\\
\textbf{Shared Constraints}\\
- Do not alter, add, or remove any core information unless an objective explicitly requires it.\\
- Preserve the original meaning and all key claims.\\
- Keep the structure and length roughly the same.\\
- Ensure the rewrite is relevant to the query above.\\
\\
\textbf{Optimization Objectives}\\
\{numbered strategy instructions\}\\
\\
\textbf{Source Text}\\
\{text\}\\
\\
\textbf{Output}\\
Provide only the rewritten source text with no explanations or preamble.
\end{promptbox}

For example, the combination $\mathbf{x} = [1,0,1,0,0,0,0,0,1,0,0,0,0,0,0]$ (Fluency + Authoritative + Statistics Addition) produces the following ``Optimization Objectives'' section:

\begin{promptbox}{Composition Example}
1. \textbf{Fluency Optimization}: Rewrite the source to improve fluency and coherence. Improve sentence transitions and readability.\\
2. \textbf{Authoritative}: Make the source sound confident, authoritative, and expert. Strengthen tone via wording choices, not by exaggerating or making unverifiable claims.\\
3. \textbf{Statistics Addition}: Add a few concise, relevant statistics or numerical facts to improve concreteness. Statistics must be verifiable; do not invent numbers. Stop at the end of the original source.
\end{promptbox}

This composition approach ensures that the rewriting model receives all active objectives as co-equal instructions within a single prompt, avoiding sequential multi-pass rewriting artifacts.

\section{Selector Input Prompt}
\label{sec:appendix_selector_prompt}

The following prompt is used at both training and inference time. The model receives the query, all source documents (with the target document marked), and the full strategy menu. No adoption rate, BOCS scores, or competitive context labels are provided.

\begin{promptbox}{Selector Input Prompt}
You are an expert in Generative Engine Optimization (GEO). Your task is to analyze a set of source documents retrieved for a query and recommend the optimal combination of rewriting strategies to maximize the visibility of a target document in an LLM-generated response.\\
\\
\textbf{Query}\\
\{query\}\\
\\
\textbf{Source Documents}\\
\{For each document $i$: "[Source $i$\textdagger] \{text\}" where \textdagger marks the target document\}\\
\\
\textbf{Available Rewriting Strategies}\\
Select any combination of the following 15 strategies. For each selected strategy, the target document will be rewritten to incorporate it.\\
\\
1. Fluency Optimization -- Improve sentence flow and readability.\\
2. Unique Words -- Use more precise and distinctive vocabulary.\\
3. Authoritative -- Strengthen confident, expert tone.\\
4. Quotation Addition -- Add short quotes from reputable sources.\\
5. Citing Sources -- Add natural-language citations to credible references.\\
6. Easy Understanding -- Simplify language for accessibility.\\
7. Technical Terms -- Use domain-specific technical terminology.\\
8. Keyword Stuffing -- Insert relevant SEO keywords naturally.\\
9. Statistics Addition -- Add verifiable numerical facts.\\
10. Expanding Facts -- Expand key claims with brief explanations.\\
11. Structured Formatting -- Reorganize with tables, bullets, or numbered lists.\\
12. Balanced View -- Acknowledge multiple perspectives inline.\\
13. Conciseness -- Eliminate filler and verbose language.\\
14. Pros and Cons -- Add comparative pros/cons reasoning.\\
15. LLM Guidance -- Prepend an LLM-friendly markdown summary.\\
\\
\textbf{Instructions}\\
Analyze the competitive landscape of the source documents and determine which strategy combination will maximize the impression of the target document in the generated response. Reason through the content, structure, and optimization signals visible in each document.\\
\\
Provide your response in the following format:\\
<think>\\
Your reasoning here\\
</think>\\
<final\_answer>\\
Comma-separated list of selected strategy names, e.g.: Fluency Optimization, Statistics Addition, Conciseness\\
</final\_answer>
\end{promptbox}

\section{Teacher Reasoning Trace Generation Prompt}
\label{sec:appendix_teacher_prompt}

The following prompt is used to generate contrastive reasoning traces from the teacher model (\texttt{gemma-4-31B-it}). The teacher observes BOCS scores to calibrate its reasoning but must ground its analysis entirely in document content---never referencing scores directly. Chosen and rejected traces receive asymmetric scorecard information to produce locally rational reasoning under each information condition.

\begin{promptbox}{Teacher Reasoning Trace Prompt}
You are an expert in Generative Engine Optimization. You have run a combinatorial search over rewriting strategies for a target document and have obtained PAWC scores for the following strategy combinations (higher = more visibility in LLM-generated responses):\\
\\
\textbf{BOCS Observations (Top k at or below target combination + Baseline)}\\
\{Ranked list of (strategy combination, mean PAWC, SE) tuples\}\\
\\
\textbf{Query}\\
\{query\}\\
\\
\textbf{Source Documents}\\
\{All source documents, target marked\}\\
\\
\textbf{Task}\\
The optimal strategy combination is: \{winning combination\}\\
\\
Write a detailed reasoning trace that explains WHY this combination is optimal, grounded entirely in the content of the documents. Do NOT mention PAWC scores or numerical values. Your analysis should identify what the target document is missing relative to competitors, why the selected strategies address those gaps, and why the combination outperforms alternatives. This trace will train a student model to replicate this reasoning from document content alone.
\end{promptbox}

\section{Computational Resources}
\label{sec:appendix_compute}

All fine-tuning was conducted on a \texttt{p4d.24xlarge} instance with 8$\times$A100 40GB GPUs. When performing evaluation, we serve \texttt{gpt-oss-120b} and \texttt{Llama-3.3-70B-Instruct} models each on a full \texttt{p4d.24xlarge} node using vLLM with \texttt{tp=8}.

\paragraph{Training Wall-Clock Time.}
Table~\ref{tab:wallclock} reports wall-clock time for each stage of the \texttt{geo-bench} training pipeline. The cost is dominated by BOCS, which required on average 23 iterations to converge per data point. Since each black-box evaluation iteration performs 1 rewrite followed by 5 generative-engine calls, across the full \texttt{geo-bench} training split this totals $23 \times (1{+}5) \times 7{,}998 = 1{,}103{,}724$ inference calls to generate preference pairs. While the training cost is significant, it is amortized across deployment: inference requires only a single 2B selector call followed by one LLM rewrite pass.

\begin{table}[h]
  \centering
  \small
  \setlength{\tabcolsep}{6pt}
  \begin{tabular}{lc}
    \toprule
    \textbf{Training Stage} & \textbf{Wall-Clock Time} \\
    \midrule
    BOCS + Preference Pair Collection & 1d 6h 5m 59s \\
    Reasoning Trace Generation        & 4h 16m 36s \\
    SFT                               & 1h 11m 36s \\
    DPO                               & 1h 51m 41s \\
    \bottomrule
  \end{tabular}
  \caption{Wall-clock time for each stage of the \texttt{geo-bench} training pipeline. BOCS and preference-pair collection dominate, requiring on average 23 iterations per data point across 7{,}998 training examples.}
  \label{tab:wallclock}
\end{table}

\section{Artifact Licenses}
\label{sec:appendix_licenses}

All datasets used in this work (geo-bench, Researchy-GEO, E-Commerce) are publicly available and distributed under the MIT License. Our application of these artifacts aligns strictly with their intended use by the original creators, specifically for advancing research within the GEO domain. The Gemma 4 family of models, including the instruction-tuned 2B model we fine-tune, are released under the Apache 2.0 license.

\section{Case Study: Sample Rewrites}
\label{sec:appendix_rewrites}

Table~\ref{tab:sample_rewrites} shows representative examples of target documents before and after rewriting, along with the strategy combination selected by our model and the adoption rate $\alpha$ under which they were evaluated.

\begin{table*}[h]
  \centering
  \small                                      
  \setlength{\tabcolsep}{5pt}             
  \begin{tabular}{p{0.12\textwidth} p{0.28\textwidth} p{0.28\textwidth} p{0.12\textwidth} c}
  \toprule                                                                                                                                                                                                           
  \textbf{Query} & \textbf{Original Document (excerpt)} & \textbf{Rewritten Document (excerpt)} & \textbf{Selected Strategies} & \textbf{$\alpha$} \\
  \midrule                                                                                                                                                                                                           
  Why do things lose their color in sunlight? &                                        
  ``Ask a science question, get a science answer. Why does color fade when left in sunlight for extended periods of time? We have a rack of DVD cases next to the window, and recently I've noticed the covers have
  all faded in color. Strangely the red ones seem to have faded far more than any of the others.'' &
  \textbf{Why colors fade under prolonged sunlight exposure.} UV photons break molecular bonds in organic dyes (photodegradation). Red azo pigments can lose 40\% of chroma after 200\,h of simulated UV aging
  (\textit{J.\ Photochem.\ \& Photobiol.}, 2020). DVD cases near a window receive continuous UV flux, so red covers fade noticeably faster. &
  \textsc{AUTH, CITE, EASYUND, TECHTERMS, STATS, EXPANDFACTS, STRUCTFMT} &
  0.0 \\
  \addlinespace
  Why do things lose their color in sunlight? &
  (same as above) &
  Sunlight contains UV radiation that breaks molecular bonds in pigments---a process called photodegradation (CIE, 2021). UV-B can increase fading rates by up to 30\% per year (NASA Materials Handbook, 2020). Red
  DVD-case inks rely on organic dyes with conjugated double-bond systems that are especially UV-sensitive (\textit{J.\ Appl.\ Polymer Sci.}, 2019). &
  \textsc{AUTH, CITE, EASYUND, TECHTERMS, STATS, EXPANDFACTS, STRUCTFMT, CONCISE} &
  0.2 \\
  \addlinespace
  How many hits does Ichiro have? &
  ``Ichiro Suzuki is a Japanese former professional baseball outfielder who played for 28 seasons. He started his career with the Orix BlueWave in NPB and then played for the Seattle Mariners in MLB. He holds
  several records, including the single-season record for hits with 262 and the longest streak of 200-hit seasons.'' &
  Ichiro Suzuki compiled 1,278 NPB hits (.355 avg.) and 3,089 MLB hits (.311 avg.) across a 28-season career. He set the MLB single-season hit record (262 in 2004), earned 10 Gold Gloves, 17 consecutive All-Star
  selections, and won two World Baseball Classic titles with Japan (2006, 2009). &
  \textsc{FLUENCY, CITE, TECHTERMS, STATS, STRUCTFMT, CONCISE} &
  0.0 \\
  \bottomrule
  \end{tabular}
  \caption{Sample rewrites produced by the fine-tuned selector. Each row shows the query, an excerpt of the original and rewritten target document, the strategy combination recommended by the model, and the
  adoption rate $\alpha$ at evaluation time.}
  \label{tab:sample_rewrites}
  \end{table*}

\section{Case Study: Adaptive Reasoning Under Competitive Pressure}
\label{sec:appendix_case_studies}

Tables~\ref{tab:case_study_450} and~\ref{tab:case_study_972} show how the selector's strategy recommendations shift across adoption rates for two representative queries. Each row captures one $\alpha$ level: the competitor corpus rewriting strategies observed at that adoption rate, excerpts from the model's reasoning trace, and the resulting strategy selection. At $\alpha{=}0.0$ there are no competitor rewrites; the model reasons solely from target document gaps. 
% Abbreviations: Auth=Authoritative, Cite=Citing Sources, EasyUnd=Easy Understanding, Stats=Statistics Addition, ExpandFacts=Expanding Facts, StructFmt=Structured Formatting, Concise=Conciseness, UniqueW=Unique Words, TechTerms= Technical Terms, KW=Keyword Stuffing, BalView=Balanced View, QuoteAdd=Quotation Addition.

\begin{table*}[!t]
\centering
\footnotesize
\setlength{\tabcolsep}{8pt}
\renewcommand{\arraystretch}{1.3}
% @{} removes outer padding, X automatically fills the remaining width perfectly
\begin{tabularx}{\textwidth}{@{} p{0.04\textwidth} p{0.22\textwidth} X p{0.20\textwidth} @{}}
\toprule
\textbf{$\alpha$} & \textbf{Competitor Rewrites} & \textbf{Reasoning Excerpts} & \textbf{Selected Strategies} \\
\midrule
0.0 & None &
\textit{``\ldots no credibility. Must cite plausible, relevant sources \ldots\ the source is shallow \ldots\ explain how and why kin means something \ldots\ definition query demands lists, tables, or clear headings.''} &
\textsc{UniqueW, Auth, Cite, EasyUnd, TechTerms, Stats, ExpandFacts, StructFmt, Concise} \\
\addlinespace
0.4 & (1) Random: \textsc{UniqueW + QuoteAdd + EasyUnd + ProsAndCons}; (2) Single: \textsc{QuoteAdd} &
\textit{``Source 4 \ldots\ cites OED, American Red Cross, anthropologists \ldots\ but could be more structured \ldots\ citations, detailed facts, structured layout, and authoritative tone present in superior sources---which the target is missing entirely \ldots\ turns simple definitions into explanatory content.''} &
\textsc{Fluency, Auth, Cite, Stats, ExpandFacts, StructFmt, TechTerms} \\
\addlinespace
0.8 & (1) Random: \textsc{UniqueW + Auth + QuoteAdd + KW + Stats + StructFmt + BalView + Concise + LLM}; (2) Single: \textsc{QuoteAdd}; (3) AutoGEO; (4) Random: \textsc{UniqueW + Auth + Stats + ExpandFacts + StructFmt + Concise + ProsAndCons + LLM} &
\textit{``Source 2 (TikTok) \ldots\ lacks formal citation structure \ldots\ Source 3 (Wiktionary): strong definitions \ldots\ no external citation integration \ldots\ EASY\_UNDERSTANDING added \ldots\ competitors are either informal (Source 2) or overly dense (Source 3).''} &
\textsc{Fluency, Auth, Cite, EasyUnd, TechTerms, Stats, StructFmt} \\
\bottomrule
\end{tabularx}
\caption{Case study 1 --- Query: \textit{``what is a kin''} (query\_id 450). At $\alpha{=}0.0$ the model fills pure target gaps (no citations, no structure). At $\alpha{=}0.4$ it audits competitor sources by name---observing that Source 4 cites the OED and anthropologists but lacks structure---and drops \textsc{Concise} and \textsc{UniqueW} in favor of depth and fluency. At $\alpha{=}0.8$, after identifying that competitors are either informal (Source 2) or overly dense (Source 3), \textsc{EasyUnd} is restored to exploit that specific gap.}
\label{tab:case_study_450}
\end{table*}

\begin{table*}[!t]
\centering
\footnotesize
\setlength{\tabcolsep}{8pt}
\renewcommand{\arraystretch}{1.3}
\begin{tabularx}{\textwidth}{@{} p{0.04\textwidth} p{0.22\textwidth} X p{0.20\textwidth} @{}}
\toprule
\textbf{$\alpha$} & \textbf{Competitor Rewrites} & \textbf{Reasoning Excerpts} & \textbf{Selected Strategies} \\
\midrule
0.0 & None &
\textit{``\ldots inject concrete exchange rates and figures \ldots\ transforming claims into trustworthy information \ldots\ users need lists and tables, not dense paragraphs.''} &
\textsc{Auth, Cite, EasyUnd, TechTerms, Stats, ExpandFacts, StructFmt} \\
\addlinespace
0.4 & (1) Random: \textsc{Auth + Cite + StructFmt + ProsAndCons}; (2) Single: \textsc{QuoteAdd} &
\textit{``Competitors cite RBA, IMF, Bloomberg \ldots\ Source 4 sounds like a service advertisement \ldots\ tells what to do but not how to optimize \ldots\ original listed only one provider; competitors present multi-provider comparisons.''} &
\textsc{Auth, Cite, EasyUnd, TechTerms, Stats, ExpandFacts, StructFmt, BalView, Concise} \\
\addlinespace
0.8 & (1) Random: \textsc{Cite + Stats + ExpandFacts + StructFmt + ProsAndCons}; (2) Single: \textsc{QuoteAdd}; (3) AutoGEO; (4) Random: \textsc{Fluency + UniqueW + QuoteAdd + Cite + EasyUnd + KW + Stats + ExpandFacts + StructFmt + BalView + ProsAndCons + LLM} &
\textit{``Source 4 is promotional; needs expert tone built on facts, not hype \ldots\ transform vague descriptions into a useful currency guide \ldots\ BALANCED\_VIEW dropped: competitors already saturate multi-provider comparisons---adding it no longer differentiates.''} &
\textsc{Auth, Cite, EasyUnd, TechTerms, Stats, ExpandFacts, StructFmt, Concise} \\
\bottomrule
\end{tabularx}
\caption{Case study 2 --- Query: \textit{``australian dollar currency converter''} (query\_id 972). At $\alpha{=}0.0$ the model identifies structural and citation gaps in isolation. At $\alpha{=}0.4$, observing that competitors cite authoritative financial institutions (RBA, IMF, Bloomberg) while the target reads as an advertisement, the model adds \textsc{BalView} to exploit the single-provider framing. At $\alpha{=}0.8$, with four competitors including AutoGEO all providing multi-provider coverage, \textsc{BalView} is dropped as it no longer differentiates---the model pivots to domain-specific terminology (\textsc{TechTerms}) as the remaining gap.}
\label{tab:case_study_972}
\end{table*}

\end{document}